\documentstyle[12pt]{article}
\begin{document}
\begin{center}
{\bf \huge Ricci  Collineations of the Bianchi Types I and III,
and Kantowski-Sachs Spacetimes }\\[1cm] U.CAMCI \footnote{e-mail :
ucamci@comu.edu.tr}$^{,\dagger}$, \.I. YAVUZ \footnote{e-mail :
iyavuz@comu.edu.tr}$^{,*}$,H. BAYSAL \footnote{e-mail :
hbaysal@comu.edu.tr}$^{,\dagger}$, \.I. TARHAN \footnote{e-mail :
tarhan\_ism@gursey.gov.tr}$^{,\ddagger}$,\\ and \.I. YILMAZ
\footnote{e-mail : iyilmaz@comu.edu.tr}$^{,\ddagger}$
\\[1cm]
$^\dagger$Department of Mathematics, Art and Science Faculty,
\d{C}anakkale Onsekiz Mart University, 17100 \d{C}anakkale, Turkey
\\[2mm] $^\ddagger$Department of Physics, Art and Science Faculty,
\d{C}anakkale Onsekiz Mart University, 17100 \d{C}anakkale, Turkey
\\[2mm] $^*$Department of Computer Science, \d{C}anakkale
Onsekiz Mart University, 17100 \d{C}anakkale, Turkey
\\[1cm]
\end{center}

\begin{abstract}

Ricci collineations of the Bianchi types I and III, and
Kantowski-Sachs space- times are classified according to their
Ricci collineation vector (RCV) field of the form (i)-(iv) one
component of $\xi^a (x^b)$ is nonzero, (v)-(x) two components of
$\xi^a (x^b)$ are nonzero, and (xi)-(xiv) three components of
$\xi^a (x^b)$ are nonzero. Their relation with isometries of the
space-times is established. In case (v), when $det(R_{ab}) = 0$,
some metrics are found under the time transformation, in which
some of these metrics are known, and the other ones new. Finally,
the family of contracted Ricci collineations (CRC) are presented.
\end{abstract}

\section{Introduction}

The general theory of relativity is described by the Einstein
field equations of the form
\begin{equation}
G_{ab}\equiv R_{ab} - \frac{1}{2} R \, g_{ab} =\kappa T_{ab}
\label{eineq}
\end{equation}
where $G_{ab}$ represents the components of Einstein tensor,
$R_{ab}$ are the components of the Ricci tensor and $T_{ab}$ being
the components of the energy-momentum (or matter) tensor. Here $R
(\equiv g^{ab} R_{ab})$ is the Ricci scalar and $\kappa$ the
Einstein gravitational constant. Since the Einstein tensor is
related, via the Einstein field equations, to the material content
of the space-time, it appears then natural to look into the
symmetries of the Einstein and energy-momentum tensors. Taking the
Lie derivative on both sides of equation (\ref{eineq}), it turns
out that the symmetries of the Einstein tensor, which is a
function of Ricci and metric tensors, and Ricci scalar, are
identically the same as the symmetries of the energy- momentum
tensor. In this paper, we restrict our attention to the Ricci
tensor symmetries.

    According to the classification of Katzin {\it et al.} \cite{KLD},
Ricci (RC) and contracted Ricci (CRC) collineations, which are the
most general symmetry transformations admitted by a given metric,
are defined by
\begin{equation}
\pounds _\xi R_{ab} = 0, \label{rc1} \\ \quad  g^{ab} \pounds _\xi
R_{ab} \label{crc} = 0,
\end{equation}
where "$\pounds _\xi$" denotes Lie derivative along $\xi$.  In a
torsion free space in coordinate basis, the RC equations can also
be written in component form as
\begin{equation}
R_{ab,c} \xi^c + R_{ac}\xi^c_{,b} + R_{cb} \xi^c_{,a} = 0,
\label{rc2}
\end{equation}
where $\xi^c$ are the components of the Ricci collineation vector
(RCV) and "$,a$" represents differentiation with respect to $x^a$.
Each member of the family of CRC symmetry mappings satisfies
$g^{ab}\pounds _\xi(T_{ab} - g_{ab} T/2) = 0$, which leads to the
following generator for a space-time admitting RC \, \cite{davis}
$$ \bigtriangledown \left[ (-g)^{1/2} \Big{(}T^a_b - \frac{1}{2}
\delta^a_b \Big{)}\xi^b \right] = \partial_a \left[ (-g)^{1/2}
R^a_b \xi^b \right]$$ where $\bigtriangledown$ and $\partial$
represent, respectively, the covariant and partial derivative
operators. Also, Davies {\it et al.} \cite{davis} have pointed out
the need for more detailed investigations of the symmetry
properties lying between isometries that leave invariant the
metric tensor under Lie transport, i.e. $\pounds _k g_{ab} = 0$,
where $k$ is a Killing vector (KV), and RCVs, as well as for more
detailed investigations of their interconnections. Space-times
admitting isometries have been widely studied by Kramer {\it et
al.} \cite{kramer}

    In recent years, much interest has been shown in the study of the various
symmetries (particularly matter and Ricci collineations) that
arise in understanding the general theory of relativity more
deeply. Green {\it et al.} \cite{green} and Nunez {\it et al.}
\cite{nunez} have considered an example of RC and the family of
CRC symmetries of Robertson-Walker metric, and they have confined
their study to symmetries generated by the vector fields of the
following form, respectively,\[ \xi = \xi^4(r,\theta,\phi,t)
\frac{\partial}{\partial t}\qquad  and \qquad \xi =
\xi^1(r,t)\frac{\partial}{\partial r} +
\xi^4(r,t)\frac{\partial}{\partial t}.\] Also, the relationship
with constants of motion between RC and family of CRC has been
indicated \cite{oliver}-\cite{KL1981}. A complete classification
of RCs was obtained for spherically symmetric static spacetimes
\cite{bokhari}, which were compared with KVs admitted by the
corresponding spacetimes. Amir {\it et al.}\cite{amir} have
investigated the relationship between the RCVs and the KVs for
these spacetimes in detail. Carot {\it et al.} \cite{carot} have
studied matter collineations, as a symmetry property of the
energy-momentum tensor Tab and also, Hall {\it et al.} \cite{hall}
have presented a discussion of Ricci and matter collineations in
space-time.

    In recently , we have discussed that RCs and family
of CRCs of Bianchi type- II, VIII, and IX spacetimes by
considering some RCVs \cite{camci}. Now, we consider that the
metric for Bianchi types I($\delta=0$) and III($\delta=-1$), and
Kantowski-Sachs ($\delta =+1$) cosmological models  is in the form
\cite{maccallum1}-\cite{lorenz},
\begin{equation}
ds^2 = dt^2 -A^2 dr^2-B^2(d\theta^2 + f^2 d\phi^2), \label{metric}
\end{equation}
which is spatially homogeneous, has shear, and has no rotation;
where $A$ and $B$  are functions of $t$ only, and for $\delta = 0,
-1, +1$, respectively,
\begin{equation}
f(\theta)= \left[
\begin{array}{lll}
\quad \theta \\ \sinh \theta \\ \, sin \theta
\end{array}
\right] \label{ffonk}
\end{equation}
(prime denotes derivative with respect to $\theta$). Further, we
observe that
\begin{equation}
f^2 \left( \frac{f'}{f} \right)' = -1 \quad  \Leftrightarrow \quad
(f \, f')' = 2f'^{\,2} -1. \label{property}
\end{equation}

   The nonvanishing Ricci tensor components of the above metric are as follows
\begin{eqnarray}
R_{11} &=& A^2 \left[ \frac{\ddot{A}}{A} + 2 \frac{\dot{A}
\dot{B}}{A\,B} \right], \nonumber \\ R_{22} &=& B^2 \left[
\frac{\ddot{B}}{B} + \frac{\dot{A} \dot{B}}{A\,B} +
\left(\frac{\dot{B}}{B} \right)^2 + \frac{\delta}{B^2} \right],
\nonumber \\ R_{33} &=& f^2 R_{22}, \nonumber
\\R_{44} &=& - \left[ \frac{\ddot{A}}{A} + 2 \frac{\ddot{B}}{B}
\right], \nonumber
\end{eqnarray}
where the dot represents derivative with respect to time. The
$\delta = 0$ Bianchi type I model clearly contains the flat
Robertson-Walker model as a particular case, but the other two are
eternally anisotropic in the sense that sectional curvatures are
never simultaneously identical.

      In this article, we investigate the symmetry properties of the
Bianchi types I and III, and Kantowski-Sachs space-times by
considering RCVs associated to the following vector fields $\xi$ :
    (i)-(iv) one component of $\xi^a (x^b)$ is nonzero,
    (v)-(x) two components of $\xi^a (x^b)$ are nonzero,
    (xi)-(xiv) three components of $\xi^a (x^b)$ are nonzero,
where $x^a = (r,\theta,\phi,t)$ for $a = 1,2,3,4$. Then, we
substitute these collineations into each of the collineation
equations. Some of these equations, in this process, are
identically satisfied, whereas the others are not and get replaced
by a set of partial differential equations to be solved for
classifying collineations. These collineations are explicitly
derived and classified completely in section $3$, while section
$2$ presents some general results and sets up the distinction
between nondegenerate and degenerate Ricci tensors. In section
$4$, the family  of CRCs will be briefly discussed. Finally, in
the last section,  we conclude with a brief summary and discussion
of results of the present work.

\section{some general results}

    Let $\xi$ be a RC, i.e. a solution of equation (\ref{rc2}) for $\xi$.
The Lie algebra of RCs may be finite or infinite dimensional.
Then, it would be interesting to know under what conditions the
Lie algebra of RCs on a space-time manifold is finite dimensional.
When studying the Lie algebra of RCs, two cases arise naturally
according to whether $R_{ab}$ is nondegenerate, or degenerate.
While the Ricci tensor must also have a noninfinite determinant,
it can be zero \cite{amir}.
     In recent paper's
of Hall et al. \cite{hall}, they have noticed some important
results the following.

   (i) If $R_{ab}$ is degenerate, then
$rank(R_{ab}) < 4$ (i.e., $det(R_{ab}) = 0$), and we cannot
guarantee the finite dimensionality of the corresponding Lie
algebra. If $R_{ab}$ is nondegenerate, then $rank(R_{ab}) = 4$,
and Lie algebra of RCs is finite-dimensional ($\leq 10$, and $\neq
9$).

  (ii) Assuming that  $\psi$ is a (smooth) scalar function
on space-time manifold, then from equation (\ref{rc2}), yields
\begin{equation}
\pounds _X R_{ab} = \pounds _{\psi\, \xi} R_{ab} = \psi_{,b}
R_{ac} \xi^c + \psi_{,a} R_{cb} \xi^c . \label{rc3}
\end{equation}
Therefore, it follows from eq. (\ref{rc3}) that $X = \psi \, \xi$
is also a RC if and only if either $\psi$ is constant on the
manifold or $\xi$ satisfies $R_{ab} \xi^b = 0$, in which case
$R_{ab}$ is necessarily degenerate (i.e., its rank is less than
$4$). That is, in the latter case, it follows that if this is the
case for $\xi \neq 0$, then the vector space of RCs on space-time
manifold is infinite dimensional.

\section{Ricci Collineations}

    For the Bianchi type I and III, and Kantowski-Sachs spacetimes, substituting
$R_{\alpha \alpha}$ ($\alpha =1,2,4$) into Eq. (\ref{rc2}) we
obtain the following RC equations :
\begin{eqnarray}
\dot{R}_{11} \xi^4 + 2 R_{11}\xi^1_{,1} = 0, \label{rceq1}\\
\dot{R}_{22}\xi^4 + 2 R_{22} \xi^2_{,2} = 0, \label{rceq2}\\
f\dot{R}_{22} \xi^4 + 2 R_{22} \Big{(}f'\xi^2 + f \xi^3_{,3}
\Big{)} = 0, \label{rceq3} \\ \dot{R}_{44} \xi^4 + 2 R_{44}
\xi^4_{,4} = 0, \label{rceq4}
\\ R_{11} \xi^1_{,2} + R_{22} \xi^2_{,1} = 0, \label{rceq5} \\ R_{11} \xi^1_{,3}
+ f^2 R_{22} \xi^3_{,1} = 0, \label{rceq6} \\ R_{11}\xi^1_{,4} +
R_{44} \xi^4_{,1} = 0, \label{rceq7}\\ R_{22} \Big{(} \xi^2_{,3} +
f^2 \xi^3_{,2} \Big{)} = 0, \label{rceq8}
\\ R_{22} \xi^2_{,4} + R_{44} \xi^4_{,2} = 0, \label{rceq9} \\
f^2 R_{22} \xi^3_{,4} + R_{44} \xi^4_{,3} = 0. \label{rceq10}
\end{eqnarray}
where commas denote the partial derivatives, and  the indices
$1,2,3$, and $4$ correspond to the variables $x, y, z$ and $t$,
respectively. Note that $R_{33}$ does not appear in the above
equations, since $R_{33} = f^2 R_{22}$ .

    The nature of solution of RC equations (\ref{rceq1})-(\ref{rceq10})
changes if one (or more) of the components of the symmetry vector
$\xi^a(x^b)$ is zero.  In the above equations
(\ref{rceq1})-(\ref{rceq10}) we will consider all subcases
(i)-(xiv) of the general RCV $\xi^a$ given above section.

    We first consider the simpler cases when one component of
$\xi^a (x^b)$ is different from zero. We now consider case (i),
$\xi^1(x^b) \neq 0$. For this case, $\xi^1$ becomes constant.  In
case (ii), $\xi^2(x^a) \neq 0$, using RC equations we obtain that
$f' R_{22} \xi^2 = 0$, that is, either $f = const.$ or $R_{22} =
0$. In the latter case, $\xi^2$ is unconstrained function of
$x^a$, and in the former case $\xi^2$ is a constant. Also, the
remaining equations become identities. In case (iii), $\xi^3(x^a)
\neq 0$, using RC equations we find that $\xi^3$ is a constant,
while the other RC equations become identities. For case (iv),
$\xi^4 (x^a) \neq 0$, it follows from RC equations that $R_{11}$
and $R_{22}$ are constants (so that $R_{33} =const. \times f^2$),
and$$\xi^4 (t) = \frac{c}{\sqrt{\mid R_{44}\mid}},$$where $c$ is
an integration constant.

    Now, we obtain the other all subcases with two and three RCV
components.

{\bf Case (v)}: $\xi = (\xi^1(x^a),\xi^2(x^a),0,0)$.

    In this case, from equation (\ref{rceq3}) we find that $R_{22}
\xi^2 = 0$, Then there are two possibilities $(a) R_{22} = 0; (b)
\xi^2 = 0$. Hence, using the other RC equations, the subcase $(b)$
is reduced to the case (i) if $R_{11} \neq 0$. In this subcase, if
$R_{11} = 0$, then $\xi^1$ is unconstrained function of $x^a$,
i.e. RCV field is $\xi = \xi^1 (x^a) \partial/{\partial r}$. In
subcase $(a)$, i.e. when $R_{22} = 0$, $\xi^2$ is unconstrained
function of $x^a$ , and $R_{11}\xi^1_{,a} = 0$ for $a=1,2,3,4$.
Thus, when $\xi^1_{,a} = 0$, then $\xi^1$ is a constant, that is
$\xi = const.\times
\partial /\partial r + \xi^2(x^a)\partial / \partial \theta$.
When $R_{11} = 0$, then $\xi^1$ is unconstrained function of $x^a$
(i.e., $\xi = \xi^1(x^a)\partial / \partial r + \xi^2(x^a)
\partial / \partial \theta$).

    Thus, using the Ricci tensor components in which the case both
$R_{11}$ and $R_{22}$ are zero, we get the following differential
equations
\begin{eqnarray}
\frac{\ddot{A}}{A} + 2 \frac{\dot{A} \dot{B}}{A\, B} = 0,
\label{casev1}
\\ \frac{\ddot{B}}{B} + \frac{\dot{A} \dot{B}}{A\, B} +
\Big{(}\frac{\dot{B}}{B} \Big{)} + \frac{\delta}{B^2} = 0.
\label{casev2}
\end{eqnarray}

    For the Bianchi type I ($\delta = 0$), under the general time transformation $dt =
h(A,B)d(new time)$  where $h$ is an arbitrary functions of $A$ and
$B$, using Eqs. (\ref{casev1}) and (\ref{casev2}), we find the
solution in the form $ A = \Big{(}a t + b\Big{)}^{m/(m+2)}, B =
A^{1/m}$, where $m$ is a constant. Thus, in this case, the metric
(\ref{metric}) becomes
\begin{equation}
ds^2 = dt^2 - \Big{(}a t + b\Big{)}^{2m/(m+2)} dr^2 - \Big{(}a t +
b\Big{)}^{2/(m+2)}\Big{(}d\theta^2 + \theta^2 d\phi^2\Big{)}.
\end{equation}
This metric is a one-parameter stiff perfect fluid LRS Bianchi
type I solutions of the Einstein field equations \cite{kramer}.

    Furthermore, from Eqs. (\ref{casev1}) and (\ref{casev2}), we find that
$$\frac{\ddot{A}}{A} + \frac{\ddot{B}}{B} + 3 \frac{\dot{A}
\dot{B}}{A\, B} + \Big{(}\frac{\dot{B}}{B}\Big{)} +
\frac{\delta}{B^2} = 0$$ which leads to
\begin{equation}
\left[ B(A\,B)^{.} \right]^{.} = - \delta A. \label{eq_casev}
\end{equation}
When we use the transformation of the time coordinate  in Eq.
(\ref{eq_casev}), we obtain the following general solution for the
Bianchi type I ($\delta = 0$) and III ($\delta = -1$), and
Kantowski-Sachs ($\delta = +1$) metrics,
\begin{equation}
\Big{(} AB \Big{)}^2 = - \delta \bar{t}^2 + b_1 \bar{t} + b_2,
\end{equation}
where $b_1$ and $b_2$ are constants. Further, under the change of
the time-coordinate by $dt=B^2d\tau$, Eqs. (\ref{casev1}) and
(\ref{casev2}) are transformed into
\begin{eqnarray}
\frac{A_{\tau \tau}}{A} = 0, \label{casev3} \\ \frac{B_{\tau
\tau}}{B} - \Big{(}\frac{B_{\tau}}{B}\Big{)}^2 + \frac{A_{\tau}
B_{\tau}}{A\,B} + \delta B^2 = 0,\label{casev4}
\end{eqnarray}
where subscript represents derivative with respect to $\tau$. Now,
from Eq. (\ref{casev3}), we get $$A(\tau) = c_1 \tau + c_2.$$ In
the case of $\delta = 0$ (i.e., Bianchi type I case), the solution
of the equation (\ref{casev4}), gives $$B(\tau) =
(c_1\tau+c_2)^{c_3/c_1},$$where $c_1, c_2$,and $c_3$ are
constants. When $\delta =\pm 1$ (Bianchi III, and Kantowski-Sachs)
and $c_1 = 0$ (i.e., $A = const.$), then for the Bianchi type III
and Kantowski-Sachs space-times, from eq. (\ref{casev4}), we
obtain the following solutions, respectively, $$B(\tau) = c_4
sec\Big{(}c_4(\tau -c_5\Big{)},$$ $$B(\tau) = c_6 / cosh(c_7
\tau),$$ where $c_4, c_5, c_6$, and $c_7$ are constants.
Therefore, using the above equalities in $dt = B^2d\tau$, we find
that $t = tan(c_4 \tau + c_5)$, and $t = (c_6^2/c_7) tanh(c_7
\tau)$, respectively.

{\bf Case (vi)}. $\xi = (\xi^1(x^a),0,\xi^3(x^a),0$).

    In this case, from RC equations, we find that $\xi^1$ and $\xi^3$
are functions of $\phi$  and $r$, respectively, and Eqs.
(\ref{rceq2}), (\ref{rceq4}), and (\ref{rceq9}) become identities
if $R_{11}$ and $R_{22}$ are not vanished. Using the remaining
equation (\ref{rceq6}),  after some algebraic manipulation, we
find
\begin{eqnarray}
\xi_{(1)} &=& \frac{\partial}{\partial r}, \qquad \xi_{(2)} =
\frac{\partial}{\partial \phi}, \nonumber
\end{eqnarray}
which are the KV fields.

{\bf Case (vii)}: $\xi = (0,\xi^2(x^a),\xi^3(x^a),0)$.

    From Eqs. (\ref{rceq2}), (\ref{rceq5}) and (\ref{rceq9}),
it follows that $\xi^2 = \xi^2(\phi)$, and from Eqs. (\ref{rceq6})
and (\ref{rceq10}), $\xi^3 = \xi^3(\theta,\phi)$. When $R_{22}
\neq 0$, after a systematic integration of the equations
(\ref{rceq3}) and (\ref{rceq7}) it is shown that RCV field is
\begin{eqnarray}
\xi = -c_1 \left( sin\phi \frac{\partial}{\partial \theta} +
\frac{f'}{f} cos\phi \frac{\partial}{\partial \phi} \right) + c_2
\left( cos\phi \frac{\partial}{\partial \theta} - \frac{f'}{f}
sin\phi \frac{\partial}{\partial \phi} \right) + c_3
\frac{\partial}{\partial \phi}. \nonumber
\end{eqnarray}
If we set the constants $c_1, c_2$, and $c_3$ into $0$ or $\pm 1$,
we find the following generators of a group $G_3$;
\begin{eqnarray}
\xi_{(1)} = sin\phi \frac{\partial}{\partial \theta} +
\frac{f'}{f} cos\phi \frac{\partial}{\partial \phi}, \qquad
\xi_{(2)} = cos\phi \frac{\partial}{\partial \theta} -\frac{f'}{f}
sin\phi \frac{\partial}{\partial \phi}, \qquad \xi_{(3)} =
\frac{\partial}{\partial \phi}. \nonumber
\end{eqnarray}
These are just the KV fields associated with spherical symmetry of
the Bianchi types I and III, and Kantowski-Sachs metrics for
$\delta = 0, -1$, and $+1$, respectively \cite{maccallum1}.

\bigskip

{\bf Case (viii)}. $\xi = (\xi^1(x^a),0,0,\xi^4(x^a))$.

    From the RC equations (\ref{rceq2}) or (\ref{rceq3}), it
follows that $\dot{R}_{22}\xi^4 = 0$; thus, there are two
possibilities: ($I$) $\dot{R}_{22} =0$, i.e., $R_{22}$ is a
constant; ($II$) $\xi^4 =0$. In the first subcase, when $R_{11}$
and $R_{44}$ are not zero, then $\xi^1$ and $\xi^4$ depend on $r$
and $t$ only. For this subcase, Eq. (\ref{rceq4}) yields
\begin{equation}
\xi^4 (r,t) = \frac{A_1 (r)}{\sqrt{\mid R_{44} \mid}}, \label{xi4}
\end{equation}
where $A_1 (r)$ is an integration function over the variable $r$.
Incorporating (\ref{xi4}) into (\ref{rceq1}), and integrating the
resulting equation with respect to $r$, we obtain
\begin{equation}
\xi^1 (r,t) = -\frac{\ddot{R}}{2 R_{11} \sqrt{\mid R_{44} \mid}}
\int{A_1 (r) dr} + A_2(t), \label{xi1}
\end{equation}
where $A_2(t)$ is an integration function. Thus, using (\ref{xi4})
and (\ref{xi1}) in eq. (\ref{rceq7}), gives
\begin{equation}
\left[ \frac{\dot{R}_{11}}{2 R_{11} \sqrt{\mid R_{44} \mid}}
\right]^{.} \int{A_1 dr} = \frac{\sqrt{\mid R_{44} \mid}}{R_{11}}
A_{1,r} + \dot{A_2},
\end{equation}
where subscript represents derivative with respect to $r$.
Differentiating the above equation relative to $r$, provided that
$A_1(r) \neq 0$, yields
\begin{equation}
\frac{A_{1,rr}}{A_1} = \frac{R_{11}}{\sqrt{\mid R_{44} \mid}}
\left[ \frac{\dot{R}_{11}}{2 R_{11} \sqrt{\mid R_{44} \mid}}
\right]^{.} = \alpha, \label{constrainteq1}
\end{equation}
where $\alpha$ is a constant of separation. In this case, there
are three possibilities: $(a) \alpha > 0$; $(b) \alpha < 0$; $(c)
\alpha = 0$. From equations (\ref{xi4})-(\ref{constrainteq1}), the
results in the case $(a)$ are given in the following form :
\begin{eqnarray}
\xi^1 &=& - \frac{\dot{R}_{11}}{2 \sqrt{\alpha} R_{11} \sqrt{\mid
R_{44} \mid}} \left[ c_1 sinh r + c_2 cosh r \right] + c3  ,
\nonumber \\ \xi^4 &=& \frac{c_1 cosh(\sqrt{\alpha} r) + c_2
sinh(\sqrt{\alpha r}}{\sqrt{\mid R_{44} \mid}}, \nonumber
\end{eqnarray}
where $c_1, c_2$ and $c_3$ are constants. The solutions for case
$(b)$ are similar to case $(a)$ but $"\alpha"$ is  replaced by
$"-\alpha"$ and hyperbolic functions by trigonometric ones.

      For subcase $(c)$, we get
\begin{equation}
\frac{\dot{R}_{11}}{2 R_{11} \sqrt{\mid R_{44} \mid}} = \beta,
\label{constrainteq1_2}
\end{equation}
where $\beta$ is an integration constant. Now, there are two
cases: ($\star$) $\beta \neq 0$; ($\dagger$) $\beta = 0$. In these
subcases, the corresponding RCVs are as follows, respectively,
$$\xi^1 = -\beta \left( c_1 \frac{r^2}{2} + c_2 r \right) - c_1
\int{\frac{\sqrt{\mid R_{44} \mid}}{R_{11}} dt} + c_3 , \quad
\xi^4 = \frac{c_1 r + c_2}{\sqrt{\mid R_{44} \mid}}$$ and $$ \xi^1
= \frac{c_1}{R_{11}} \int{\sqrt{\mid R_{44} \mid} dt} + c_2 ,
\quad \xi^4 = \frac{c_1 r + c_2}{\sqrt{\mid R_{44} \mid}}, \quad
(R_{11} = const.)$$where $c_1, c_2, c_3$ are constants. The
remaining possible case is that $R_{11}$ and $R_{44}$ do not
satisfy equation (\ref{constrainteq1}). In this case $A_1 = 0$ and
resulting RCV is same as the case (i). Also, in the second subcase
($II$), RCVs are reduced the case (i).
\begin{eqnarray}
\xi_{(1)} &=& \frac{\partial}{\partial \theta}, \nonumber \\
\xi_{(2)} &=& \frac{1}{\sqrt{\mid R_{44} \mid}}
\frac{\partial}{\partial t} . \nonumber
\end{eqnarray}

\newpage

{\bf Case (ix)}. $\xi = (0,\xi^2(x^a),0,\xi^4(x^a))$.

    In this case, from equation (8), we find that there are two
possibilities: ($I$) $\dot{R}_{11} = 0$, i.e. $R_{11} =$ const.;
($II$) $\xi^4 = 0$. The latter case is reduced to the case (ii).
In subcase ($ix.I$), from eqs. (\ref{rceq5}), (\ref{rceq7}),
(\ref{rceq8}), and (\ref{rceq10}), we obtain that $\xi^2$ and
$\xi^4$ emerge as arbitrary functions of $\theta$ and $t$. Thus,
eq. (\ref{rceq4}) yields
\begin{equation}
\xi^4 (\theta,t) = \frac{B_1 (\theta)}{\sqrt{\mid R_{44} \mid}},
\end{equation}
where $B_1(\theta)$ is an integration function, and $R_{44} \neq
0$. Subtracting Eq. (\ref{rceq2}) from Eq. (\ref{rceq3}), and
integrating the resulting equation with respect to $\theta$, gives
\begin{equation}
\xi^2(\theta ,t) = f B_2(t),
\end{equation}
where $f$ take values $\theta, sinh \theta, sin \theta$ for the
Bianchi type $I$ and $III$, and Kantowski-Sachs space-times,
respectively, and $B_2(t)$ is an integration function. Thus, using
eq. (\ref{property}), it follows that
\begin{equation}
\frac{\dot{R}_{22}}{2 R_{22} \sqrt{\mid R_{44} \mid}}\frac{1}{B_2
(t)} = - \frac{f'}{B_1(\theta)} = \gamma, \label{constrainteq2}
\end{equation}
where $\gamma$ is a separation constant which is different from
zero but it can take the value of $1$ without losing of
generality. Thus, from the constraint (\ref{constrainteq2}), we
are left with
\begin{equation}
B_1(\theta) = -f', \qquad B_2(t) = \frac{\dot{R}_{22}}{2 R_{22}
\sqrt{\mid R_{44} \mid}}, \quad
\end{equation}
Therefore, the components $\xi^2$ and $\xi^4$ are
\begin{equation}
\xi^2 = \frac{\dot{R}_{22}}{2 R_{22} \sqrt{\mid R_{44} \mid}},
\quad \xi^4 = - \frac{f'}{\sqrt{\mid R_{44} \mid}}.
\end{equation}
Thus, using (\ref{ffonk}), it follows from eq.(\ref{rceq9}) that
the other constraint equation is
\begin{equation}
\frac{R_{22}}{\sqrt{\mid R_{44} \mid}} \left[
\frac{\dot{R}_{22}}{2 R_{22} \sqrt{\mid R_{44} \mid}} \right]^{\bf
.} = \frac{f''}{f} = -\delta, \label{constrainteq3}
\end{equation}
where $\delta$ take values of $0$ (Bianchi I), $-1$ (Bianchi III),
or $+1$ (Kantowski-Sachs). In the case of $\delta = 0$ (Bianchi
I), from the constraint (\ref{constrainteq3}), we get
\begin{equation}
\frac{\dot{R}_{22}}{2 R_{22} \sqrt{\mid R_{44} \mid}} = \eta,
\end{equation}
where $\eta$ is an integration constant. In the case of $\eta =
0$, it reduce to the case (iv). If $\eta \neq 0$, then the RCVs
are
\[ \xi = \eta \theta \frac{\partial}{\partial \theta} +
\frac{1}{\sqrt{\mid R_{44} \mid}} \frac{\partial}{\partial t}. \]

\bigskip

{\bf Case (x)}. $\xi = (0,0,\xi^3(x^a),\xi^4(x^a))$.

    In this case, it follows from Eqs. (\ref{rceq1}) and (\ref{rceq2})
that $R_{11}$ and $R_{22}$ are constants (when $\xi^4 = 0$, this
case is reduced to the case (ii)). From Eqs. (\ref{rceq6}) and
(\ref{rceq8}), and using the last result in Eq. (\ref{rceq3}), we
find that $\xi^3$ is a function of $t$ only, while from Eqs.
(\ref{rceq7}) and (\ref{rceq9}), we have that $\xi^4$ is functions
of $\phi$ and $t$ . Also, Eq. (\ref{rceq4}) gives
\begin{equation}
\xi^4 (\theta,t) = \frac{D(\phi)}{\sqrt{\mid R_{44} \mid}},
\end{equation}
where $D(\phi)$ is a function of  integration with respect to $t$.
Plugging this equation into Eq. (\ref{rceq10}) and
differentiating, we obtain the following RCVs \[ \xi_{(1)} =
\frac{\partial}{\partial \theta}, \qquad \xi_{(2)} =
\frac{1}{\sqrt{\mid R_{44} \mid}} \frac{\partial}{\partial t}. \]

{\bf Case (xi)}. $\xi = (\xi^1(x^a),\xi^2(x^a),\xi^3(x^a),0)$.

    In this case, from eqs. (\ref{rceq1}), (\ref{rceq2}), (\ref{rceq7}),
and (\ref{rceq10}), it follows that $\xi^1 = \xi^1(\theta,\phi),
\,\, \xi^2 = \xi^2(r,\theta)$, and $\xi^3 = \xi^3(r,\theta,\phi)$.
Then, from eq. (\ref{rceq5}), yields
\begin{eqnarray}
\xi^1 &=& A_1(\phi) \theta  + A_2(\phi), \nonumber \\ \xi^2 &=& -
a A_1(\phi) r + A_3(\phi), \nonumber \\ R_{11} &=& a R_{22},
\nonumber
\end{eqnarray}
where $a$ is a constant; $A_1, A_2$, and $A_3$ are integration
constants. Using above mentioned results in eq. (\ref{rceq6}), we
find
\[ \xi^3 = \frac{a r}{f^2} \left( A_{1,\phi} \theta + A_{2,\phi}
\right) + B_1(\theta,\phi) \] where $B_1(\theta,\phi)$ is a
constant of integration. If $a = 0$, i.e. $R_{11} = 0$, then
$\xi^1$ is unconstrained function of $x^a$, and the other
components are the same as in the case (vii). Clearly the vector
space of RCVs is infinite dimensional in the latter case. When $a
\neq 0$ and $\delta = 0$ (Bianchi I), then substituting $\xi^2$
and $\xi^3$ into eq. (\ref{rceq3}), after some algebra, we obtain
that
\begin{eqnarray}
A_1 & = & c_1 cos\phi + c_2 sin\phi, \nonumber  \\ A_2 &=& c_3
\phi + c_4, \nonumber \\ B_1 &=& -\frac{1}{\theta} \int{A_3 d\phi}
+ A_4(\theta), \nonumber
\end{eqnarray}
where $c_i \, (i = 1,2,3,4)$, are constants, and $A_4$ is a new
integration function. Thus, we find from eq. (\ref{rceq8}) that
$c_3$ vanishes, and also $A_3$ and $A_4$ take the following form
\begin{eqnarray}
A_3 &=& c_5 cos\phi + c_6 sin\phi, \nonumber \\ A_4 &=& c_7,
\nonumber
\end{eqnarray}
where $c_5, c_6$, and $c_7$ are constants. Therefore, relabelling
$c_4 \leftrightarrow c_3, c_5 \leftrightarrow c_4, c_6
\leftrightarrow c_5$, and $c_7 \leftrightarrow c_6$, and setting
these parameters into $0$ or $\pm 1$,  we obtain $R_{11} = a
R_{22}$ and six collineations
\[ \xi_{(3)} = \frac{\partial}{\partial \phi}, \quad  \xi_{(4)} =
\theta sin\phi \frac{\partial}{\partial r}  - a r \xi_{(1)}, \quad
\xi_{(5)} = \theta cos\phi \frac{\partial}{\partial r} - a r
\xi_{(2)}, \quad \xi_{(6)} = \frac{\partial}{\partial r}, \] where
$a$ is a nonzero constant, and $\xi_{(1)}$ and $\xi_{(2)}$ are
given in the case (vii). The Lie algebra of RCVs is spanned by
these vectors. The vectors $\xi_{(1)} , \xi_{(2)}$, and
$\xi_{(3)}$ correspond to the KV fields associated with spherical
symmetry, while the other ones are the proper RCVs of the Bianchi
type I space-time. The nonvanishing commutators are given by the
following
\begin{eqnarray}
\left[ \xi_{(1)},\xi_{(3)} \right] &=& -\xi_{(2)}, \quad \left[
\xi_{(1)},\xi_{(4)} \right] = \left[ \xi_{(2)},\xi_{(5)} \right] =
\xi_{(6)}, \nonumber \\ \left[ \xi_{(2)},\xi_{(3)} \right] &=&
\xi_{(1)}, \quad \,\,\, \left[ \xi_{(2)},\xi_{(6)} \right] =
\left[ \xi_{(3)},\xi_{(4)} \right] = \xi_{(5)}, \quad \left[
\xi_{(3)},\xi_{(5)} \right] = -\xi_{(4)}, \nonumber \\ \left[
\xi_{(4)},\xi_{(5)} \right] &=& a \xi_{(3)}, \quad \left[
\xi_{(4)},\xi_{(6)} \right] = a \xi_{(1)}, \quad \left[
\xi_{(5)},\xi_{(6)} \right] = a \xi_{(2)}, \nonumber
\end{eqnarray}

    If $a \neq 0$ and $\delta \neq 0$ (for Bianchi III and
Kantowski-Sachs), then $A_1$ vanishes, and $A_2, A_3$, and $A_4$
are same as the above one. Thus, three of the four collineations
remain the same as in case (vii), whereas the other is $\xi_{(4)}
=
\partial / \partial r$. These are nonproper RCVs and  the generators
of the group $G_4$.

\bigskip

{\bf Case (xii)}. $\xi = (\xi^1(x^a),0,\xi^3(x^a),\xi^4(x^a))$.

    In this case, after a serial algebra, it follows from the
resulting RC equations that the constraints do appear in a similar
way as in case (viii), such as the eqs. (\ref{constrainteq1}) and
(\ref{constrainteq1_2}), and hence the components $\xi^1$ and
$\xi^4$ keep the forms as in case (viii), and whereas $\xi^3$
becomes a constant.

\bigskip

{\bf Case (xiii)}. $\xi = (\xi^1(x^a),\xi^2(x^a),0,\xi^4(x^a))$.

    In this case, using the RC eqs. (\ref{rceq6}), (\ref{rceq8}), and
(\ref{rceq10}), then the components $\xi^1, \xi^2$, and $\xi^4$ of
RCV emerge as an arbitrary functions of $r, \theta$, and $t$. Then
from eq. (\ref{rceq4}) yields \[ \xi^4 = \frac{A_1
(r,\theta)}{\sqrt{\mid R_{44} \mid}}, \] where $A_1(r,\theta)$ is
an integration function. Putting this component into eq.
(\ref{rceq3}), gives \[ \xi^2 = - \frac{\dot{R}_{22}}{2 R_{22}
\sqrt{\mid R_{44} \mid}} \frac{A_1 f}{f'}, \] Thus, inserting
$\xi^2$ and $\xi^4$ into eq. (\ref{rceq2}), and integrating, we
obtain
\begin{equation}
A_1 = f' B_1 (r), \label{a1}
\end{equation}
regardless of whether $\dot{R}_{22} = 0$  or not. Here, $B_1(r)$
is a function of integration. Substituting $\xi^4$ into
eq.(\ref{rceq1}), and using (\ref{a1}), we find that \[ \xi^1 =
-\frac{\dot{R}_{11}}{2 R_{11} \sqrt{\mid R_{44} \mid}} \int{B_1
(r) d r} + A_2(\theta,t), \] where $A_2$ is an arbitrary function
of the integration over the variable $r$. Then, using $\xi^2$ and
$\xi^4$ in (\ref{rceq9}), $\xi^1$ and $\xi^4$ in (\ref{rceq5}),
and differentiating, we obtain the following constraints,
respectively,
\[
\frac{R_{22}}{\sqrt{\mid R_{44} \mid}} \left[
\frac{\dot{R}_{22}}{2 R_{22} \sqrt{\mid R_{44} \mid}} \right]^{.}
= \frac{f''}{f} = -\delta,
\]
\[
\frac{R_{11}}{\sqrt{\mid R_{44} \mid}} \left[
\frac{\dot{R}_{11}}{2 R_{11} \sqrt{\mid R_{44} \mid}} \right]^{.}
= \frac{B_{1,rr}}{B_1} = \alpha,
\]
where $\delta$ take values of  $0, \pm 1$ for Bianchi type I and
III, and Kantowski-Sachs space-times and $\alpha$ is a separation
constant. Thus, there are six different cases: A) $\alpha >0$ and
$\delta \neq 0$, B)$\alpha < 0$ and $\delta \neq 0$, C)$\alpha =
0$ and $\delta \neq 0$, D)$\alpha > 0$ and $\delta = 0$, E)$\alpha
< 0$ and $\delta = 0$, F)$\alpha = 0 = \delta$.

    In subcase A), one finds $B_1 (r) = c_1 cosh r + c_2 sinh r$, and
substituting this result  into $\xi^1$ and $\xi^2$, integrating
and plugging them into (\ref{rceq5}), after some algebra, we find
\begin{eqnarray}
\xi^1 &=& -\frac{f' \dot{R}_{11}}{2 \sqrt{\alpha} R_{11}
\sqrt{\mid R_{44}\mid}} \left[ c_1 sinh\left(\sqrt{\alpha} r
\right) + c_2 cosh\left(\sqrt{\alpha} r\right) \right] + c_3,
\nonumber
\\ \xi^2 &=& -\frac{f \dot{R}_{11}}{2 \sqrt{\alpha} R_{11} \sqrt{\mid
R_{44}\mid}} \left[ c_1 cosh\left(\sqrt{\alpha} r \right) + c_2
sinh\left(\sqrt{\alpha} r\right) \right], \nonumber
\\ \xi^4 &=& \frac{f'}{\sqrt{\mid R_{44} \mid}} \left[ c_1
cosh\left(\sqrt{\alpha} r\right) + c_2 sinh\left(\sqrt{\alpha}
r\right) \right], \nonumber \\ R_{22} &=& \frac{\delta}{\alpha}
R_{11} + a, \quad  a = const. \nonumber
\end{eqnarray}
where $\delta$ take values of $\pm 1$ only. In subcase B), the
solutions are similar to the subcase A) but "$\alpha$" is replaced
by "$-\alpha$ and hyperbolic functions by trigonometric ones. In
subcase C), we get \[ B_1 (r) = d_1 r + d_2, \qquad
\frac{\dot{R}_{11}}{2 R_{11} \sqrt{\mid R_{44}\mid}} = \beta
\] where $d_1 , d_2$ are arbitrary constants, and $\beta$ is a
constant of integration. Thus, in this subcase, there are also two
possibilities : a)$\beta \neq 0$, b)$\beta = 0$. First one is
reduced to the case (i). In latter one, RCVs and the restrictions
are as follows
\[
\xi = \left( \frac{d_1 \dot{R}_{22}}{2 R_{11} \sqrt{\mid R_{44}
\mid}} \int{f d\theta} + d_3 \right) \frac{\partial}{\partial r} -
\frac{f \dot{R}_{22} (d_1 r + d_2)}{2 R_{22} \sqrt{\mid R_{44}
\mid}} \frac{\partial}{\partial \theta} + \frac{f' (d_1 r +
d_2)}{\sqrt{\mid R_{44} \mid}} \frac{\partial}{\partial t},\]
\begin{eqnarray}
\frac{\dot{R}_{22}}{\sqrt{\mid R_{44}\mid}} &=& -\delta
\int{\sqrt{\mid R_{44} \mid} d t} + const. \nonumber \\ \left(
R_{22} -1\right)\sqrt{\mid R_{44} \mid} &=& const.,\nonumber \\
R_{11} &=& const., \nonumber
\end{eqnarray}
In subcase D) and E), it follows from RC equations that $\xi^2$ is
a constant, i.e.; this case is reduced to the case (i). Finally,
in the last subcase F), i.e. when $f(\theta) = \theta$-Bianchi I,
$B_1(r) = d_1 r + d_2, \quad \beta = \dot{R}_{11}/2R_{11}$, and
$\lambda = \dot{R}_{22}/2R_{22}$, it follows from eq.(\ref{rceq5})
that $\lambda R_{22} = 0$. Therefore, if $\lambda \neq 0 \neq
\beta, R_{22} = 0$, then we get
\begin{eqnarray}
\xi^1 &=& -\beta \left( d_1 \frac{r^2}{2} + d_2 r \right) + d_1
\int{\frac{\sqrt{\mid R_{44} \mid}}{R_{11}}} + d_3, \nonumber
\\ \xi^2 &=& -\lambda \theta (d_1 r + d_2), \nonumber \\
\xi^4 &=& \frac{d_1 r + d_2}{\sqrt{\mid R_{44}\mid}}. \nonumber
\end{eqnarray}
If  $\lambda = 0,\, \beta \neq 0 \neq R_{22}$, then this case
reduces to the subcase (viii.I.c.*) and it follows $R_{22} =
const.$ If $\lambda = 0 = \beta, R_{22} \neq 0$, then it reduces
to the subcase (viii.I.c.$\dagger$), and follows that  $R_{11}$
and $R_{22}$ are constants.

{\bf Case (xiv)}. $\xi = (0,\xi^2(x^a),\xi^3(x^a),\xi^4(x^a))$.

    In this case, after some algebra, it follows from RC equations
that $\xi^4$ vanishes, and $R_{11}$ becomes constant. Therefore,
this case is reduced to the case (vii).

\section{Family of Contracted Ricci Collineations}

    For the Bianchi types I and III, and Kantowski-Sachs space-times,
the family of CRC equation (2) takes the form
\begin{eqnarray}
\frac{R_{11}}{A^2}\xi^1_{,1} + \frac{R_{22}}{B^2} \left(
\xi^2_{,2} + \frac{f'}{f}\xi^2 + \xi^3_{,3} \right) + \left(
\frac{\dot{R}_{11}}{A^2} + 2 \frac{\dot{R}_{22}}{B^2}
-\dot{R}_{44} \right) \xi^4 - R_{44} \xi^4_{,4} = 0. \label{crc}
\end{eqnarray}
However, it is not possible to find a solution to (39) without
imposing some additional restrictions either on the metric tensor,
on the Ricci tensor, or on the vector fields $\xi^a$.

    An example of proper the family of CRC is obtained by setting
each of the components of $\xi^a(x^b)$ to zero. Therefore, using
the cases (i) and (iv) of the above section in eq. (\ref{crc}), we
obtain the following equations, respectively,
\begin{eqnarray}
\frac{R_{11}}{A^2} \xi^1_{,1} = 0, \label{crc1} \\
\frac{R_{22}}{B^2} \left( \xi^2_{,2} + \frac{f'}{f} \xi^2 \right)
= 0, \label{crc2}
\\ \frac{R_{22}}{B^2} \xi^3_{,3} = 0, \label{crc3} \\ \left(
\frac{\dot{R}_{11}}{A^2} + 2 \frac{\dot{R}_{22}}{B^2} -
\dot{R}_{44} \right) \xi^4 - R_{44} \xi^4_{,4} = 0. \label{crc4}
\end{eqnarray}
Now, for the case (i), using Eq. (\ref{crc1}), we get the
possibilities: $(a.1) R_{11} = 0, (b.1) \xi^1_{,1} = 0$. In the
first subcase, $\xi^1$ is unconstrained, while in the second one,
$\xi^1 = \xi^1(\theta,\phi,t)$. In case (ii), using Eq.
(\ref{crc2}), we find that $(a.2) R_{22} = 0, (b.2) \xi^2_{,2} +
(f'/f) \xi^2 = 0$. Therefore, in subcase $(a.2)$, $\xi^2$ is
unconstrained, and in subcase $(b.2)$, we obtain \[  \xi =
F(r,\phi,t) f \partial /\partial \theta \] where $F(r,\phi,t)$ is
an integration constant, and $f$ take values $\theta, sin\theta$,
and $sinh\theta$ for Bianchi I ($\delta=0$), Bianchi III
($\delta=1$), and Kantowski-Sachs ($\delta=-1$), respectively. In
case (iii), from Eq.(\ref{crc3}), we have two different cases :
$(a.3) R_{22} = 0, (b.3) \xi^3_{,3} = 0$. Thus, in each subcases,
we get that $\xi^3$ is unconstrained, and $\xi^3=
\xi^3(r,\theta,t)$, respectively.

    In case (iv), if we set $R_{11}$ and $R_{22}$ to constants, then,
using Eq. (\ref{crc4}), we get \[ \xi^4 =
\frac{G(r,\theta,\phi)}{\sqrt{\mid R_{44} \mid}}, \] where
$G(r,\theta,\phi)$ is an integration constant. In the other cases,
it is difficult to solve the equation (\ref{crc}).

\section{Conclusion}

    In the present paper, the Bianchi types I and III, and
Kantowski-Sachs space- times are classified according to their
RCVs. In cases (i)-(iii), RCVs are coincide with KVs, and then it
follows that case (i) represents a translation along radial $r$
direction, and the other cases (ii) and (iii) represent a rotation
around $\theta$ and $\phi$ directions, respectively. The situation
arising when any of $R_{ii}$ for $i = 1,2,4$ vanishes is identical
to the fact that $R_{ab}$ is degenerate, i.e. $det(R_{ab}) = 0$,
and the corresponding Lie algebra of RCVs is infinite dimensional.
Further, in cases (ii) and (v), from (\ref{rc3}), it is shown that
$\psi X$ is also a RC, because of $R_{22} \xi^2 = 0$. Also, in
case (v), we have obtained that RCVs and KVs are identical if
$R_{ab}$ is nondegenerate, i.e. $det(R_{ab}) \neq 0$, and RCVs are
nontrivial (i.e., nonproper) if $R_{ab}$ is degenerate. It should
be noticed that (iv), (viii), (ix), (x), (xii), (xiii), and (xiv)
types of symmetry vector $\xi^a$ containing the component $\xi^4$
would already lead to proper RCV, and the cases (vi) and (vii)
which are not containing $\xi^4$ are nonproper for this
cosmological model. In case (xi), when $\delta = 0$ (Bianchi type
I), some of  the RCVs are proper but when $\delta = -1$ (Bianchi
type III and Kantowski-Sachs), then RCVs are nonproper. As pointed
out Hall {\it et al.} \cite{hall}, in classification of the
Bianchi types I and III, and Kantowski-Sachs space-times according
to their RCVs, we obtain the following results from the discussion
so far:  (a) For this cosmological model the vector space of RCVs
is infinite dimensional if $R_{ab}$ is degenerate, and is finite
dimensional if $R_{ab}$ is nondegenerate. (b) It may properly
contain the Killing or homothetic Lie algebra. For a proper
Einstein space, the vector space of RCVs coincides with the
Killing algebra. (c) Every KV is a RCV for the Bianchi types I and
III, and Kantowski-Sachs space-times, but the converse is not
true. Also, in case (v), some metrics are found under the time
transformation, in which some of them (the cases with $\delta =
0$-Bianchi I) are known \cite{kramer}, and the others new.
Further, we note that the cosmological metrics with two RCVs are
important because of the cylindrical and plane metrics, and many
of the Bianchi metrics, are special cases of $G_2$ cosmologies
\cite{lorenz}. Finally, in section $4$, for the cases (i)-(iv), we
find some proper family of CRC vector.

    In this paper, the RC equations (\ref{rceq1})-(\ref{rceq10})
have been obtained by assuming that $R_{ii}, \, (i = 1,2,4)$  do
not vanish. However, one can provide a classification of the RCs
according to the nature of the Ricci tensor. This case will be the
subject of an another article.

\newpage

\bigskip

{\bf Acknowledgments}

\bigskip

One of us, U. Camc{\i }, would like to thank Dr.G.S. Hall
(University of Aberdeen) for sending several reprints and
preprints of his work, and also express his profound gratitude to
the Feza G\"ursey Institute for the hospitality. This work was
financially supported by Turkish Scientific Research Council
(T\"UB\.ITAK) under Project Number TBAG-1674.

\end{document}